\documentstyle[12pt,preprint]{aastex}

\begin{document}

\title{Cataclysmic Variables: An Empirical Angular Momentum Loss
Prescription From Open Cluster Data.}
\author{N. Andronov, M. Pinsonneault, A. Sills\altaffilmark{1}}
\affil{Ohio State University, Department of Astronomy, Columbus, OH 43210}
\affil{E-mail: andronov, pinsono, asills@astronomy.ohio-state.edu}
\altaffiltext{1}{current address: Department of Physics and Astronomy,
McMaster University, 1280 Main St. W., Hamilton, ON, L8S 4M1, Canada}

\begin{abstract}
We apply the angular momentum loss rates inferred from open cluster
stars to the evolution of cataclysmic variables (CVs). We show that
the angular momentum prescriptions used in earlier CV studies are
inconsistent with the measured rotation data in open clusters. The
timescale for angular momentum loss ($\dot{J}$) above the fully
convective boundary is $\sim$2 orders of magnitude longer than
inferred from the older model, and the observed angular momentum loss
properties show no evidence for a change in a behavior at the fully
convective boundary. This (1) provides evidence against the hypothesis
that the period gap is caused by an abrupt change in the angular
momentum loss law when secondary becomes fully convective, (2) implies
that evolution of CV is much longer than it was thought, comparable to
a Hubble time; for the same reason, it will be more difficult to
produce CVs from the products of CE evolution and implies much lower
space density of CVs, (3) is consistent with the observed period
minimum (1.3 hours) contrary to the minimum predicted by the case when
only angular momentum loss due to gravitational radiation works (1.1
hours).

We introduce a method to infer time-averaged mass accretion rate and derive mass-period relation for different 
evolutionary states of the secondary.
The mass-period relationship is more consistent with evolved 
secondaries than with unevolved secondaries above the period 
gap.  Implications for the CV period gap are discussed, including the 
possibility that two populations of secondaries could produce the gap.

\end{abstract}

\section{Introduction}
Cataclysmic Variable stars (CVs) are mass-transferring binary systems
with orbital periods between 1.3 and 10 hours (see Patterson 1984;
Warner 1995 for reviews).  The primary in a CV is a white dwarf, and
the secondary is a low mass main sequence star (see Smith \& Dhillon
1999) which is overfilling its Roche lobe and transferring mass onto
the primary.  The evolution of a CV is driven by angular momentum
loss; it has long been known that low mass stars lose angular momentum
from a magnetic wind (e.g. Weber \& Davis 1967).  This angular
momentum is removed from the orbit in a synchronized binary system,
which causes the orbit to decay.  Gravitational radiation is a second
angular momentum loss mechanism, which is important at short
periods. Mass accretion reduces the moment of inertia of the system,
and these two effects together will determine the time evolution of CV
systems. In general, a given CV will evolve from a higher mass
secondary with a longer period to a lower mass secondary with a
shorter period.

It is the purpose of this paper to connect two different areas of
astrophysics, the study of the angular momentum evolution of low mass
stars and the study of CVs.  Following the work of Basri (1987), the
angular momentum loss properties of secondaries in CVs are typically
assumed to be the same as those for single stars or detached binary
stars.  There has been a dramatic increase in the amount and quality
of rotation data available for low mass stars in open clusters and the
field, from the important early work of Stauffer \& Hartmann (1987) to
the present (see Stauffer 1997, Krishnamurthi et al. 1997, Reid \&
Mahoney 2000 for reviews). However, many theoretical studies of CVs
use angular momentum loss rates (e.g. Rappaport, Verbunt \& Joss 1983)
that precede these data. We will show that an empirical angular
momentum loss law as a function of rotation rate and mass both poses a
challenge to our understanding of some of the important ingredients in
the study of CVs and also presents an opportunity to learn about some
crucial components by reducing the number of degrees of theoretical
freedom.

The study of CVs is a rich and complex one. Despite much
work, there some significant unresolved problems in the field:\\

1. Classification of CVs. There are several observationally distinct
classes of CVs. Nova like systems (NLs) show steady emission.  Dwarf
Novas are known for outbursts - sudden increases of their visual
brightness by 2-5 magnitudes for periods of a few days (see Verbunt
1997, Warner 1995, Patterson 1984). Magnetic CVs (AM Her systems or
mCVs) are discovered by their strong X-ray fluxes. VY Scl are the most
mysterious variables - they spend most of their time at a relatively
bright level, sometimes showing declining level of brightness.
Verbunt (1997) and Hellier \& Naylor (1998) have argued that they
should be categorized either as NLs or DNs.  The physical origin of
these differences is not understood.\\

2 The period distribution of CVs and the origin of the period
gap. There are only a few CVs observed with periods between 2 and 3
hours, while there are many systems above and below this period
range. The lack of CVs in the 2-3 hour period range seems to be
statistically significant (see Verbunt 1997, Hellier \& Naylor 2000.)
In Figure 1 we show the distribution of cataclysmic variables as a
function of orbital period; the data is taken from Clemens, Reid \&
Gizis (1998).  A successful theory of CVs must explain the shape of
this distribution and the origin of the period gap.\\

\begin{figure}[t]
\plotone{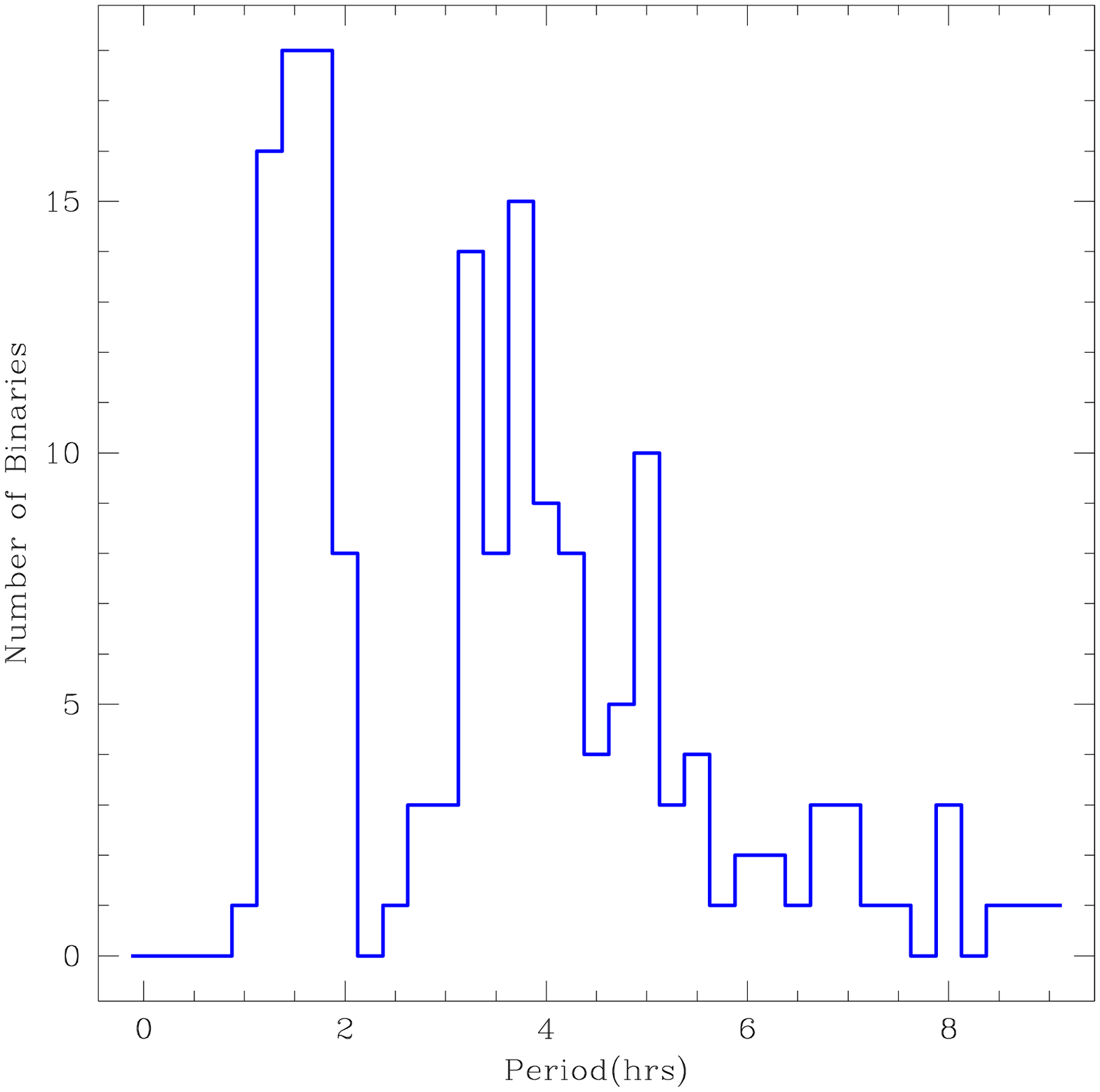}
\caption{\tiny Orbital period distribution of CVs from  Clemens, Reid \&
Gizis(1998). }
\end{figure}

The origin of the different classes of CVs necessarily involves a detailed
consideration of the physics of the accretion disk and its interaction
with the stars.  This is beyond the scope of the current paper.  It
is, however, possible to study the existence and origin of the period
gap without knowledge of the detailed properties of accretion disks.

A variety of models have been proposed attempting to explain the lack
of systems in the period range between 2 and 3 hours (period
gap). Almost all of them are different implementations of the
suggestion of Robinson et al. (1981) that some
mechanism forces secondaries to suddenly shrink when they
reach a mass of about 0.3 $M_\odot$.  Due to the correlation between the
mass of the secondary and the orbital period, this happens at about at
3 hours for typical white dwarf masses (see section 3.4).  The mass
transfer then stops until the secondary touches its Roche lobe again,
reestablishing the contact and accretion. Robinson et al. noted that this is
the characteristic mass where the secondary becomes fully convective.
Theorists have therefore focused on stellar properties that might
change at the fully convective boundary.

D'Antona and Mazzitelli (1982) proposed that sudden mixing of $^3 He$
when the star becomes fully convective could cause star to
shrink. Later calculations, however, showed that this effect is
negligible (McDermott \& Taam 1989.)  Recent stellar evolution models
do not predict any change of mass-radius relation at the point where
the star becomes fully convective (see table 1).

Rappaport, Verbunt \& Joss (1983) proposed the ``disrupted angular
momentum loss'' model.  The most recent description of this model can
be found in Howell, Nelson \& Rappaport (2000).  The timescale for
mass accretion is governed by the timescale for angular momentum loss.
The Rappaport et al. angular momentum loss timescale for a magnetic
stellar wind was shorter than the Kelvin-Helmholtz timescale.  The
secondary star, being out of thermal equilibrium, could then have a
greater radius than a normal main sequence star of the same mass.

Stellar magnetic fields are generated by a dynamo mechanism, and there
are plausible theoretical grounds for believing that stellar magnetic
field properties could be different in fully convective stars than in
stars with radiative cores.  As the total stellar mass decreases, the
depth of the outer convection zone increases; at very low masses
(
solar dynamo is thought to be anchored at the interface between the
radiative interior and the convective envelope; this mechanism will no
longer operate in a fully convective star, which requires a
distributed dynamo mechanism (Durney \& Latour 1978).  McDermot \&
Taam (1989) claimed that the net effect would be a drastic reduction
in angular momentum loss when the secondary became fully convective.
With a fully convective secondary, the timescale for angular
momentum loss increases dramatically, the secondary shrinks
back to its normal radius, and the star is not visible as a CV until
gravitational radiation could bring the stars close enough together to
cause the secondary to once again overfill its Roche lobe.  This model
is testable using stellar rotation rates in open clusters with a range
of ages and masses.  

In this paper, we will show that the empirical timescale for angular
momentum loss is much longer that predicted by early models of
$\dot{J}$, and that the secondary star will therefore be in thermal
equilibrium above and in the period range where the gap is found.
Second, the stellar transition between the two dynamo mechanisms
appears to be smooth rather than abrupt, suggesting that something
other than a phase transition in the angular momentum loss rate must
be used to explain the origin of the period gap.

Potential problems with the disrupted magnetic braking model have been
discussed in the literature (for example, Clemens et
al. 1998 noted the conflict with rotation and activity data in low
mass stars.)  The assumption that magnetic braking is the only angular
momentum loss mechanism for short period CVs also creates some difficulties
(Patterson 1998.)  Patterson concluded that if the orbital evolution
of CVs at low period were driven only by gravitation radiation there
would be two serious problems.

First, the predicted period minimum would be at 1.1 hours instead of
the observed cutoff at 1.3 hours.  Second, the angular momentum loss
rates would be very low for short period systems, which implies a low mass
accretion rate for the systems near the minimum period.  There would
then be a large number of CVs observed at or near the minimum period,
which is in contradiction with the observations.  In order to solve
these problems Patterson suggested that
there should be some mechanism for destroying CVs before they reach
the predicted very high space density.  This could be possible if
secondaries in CVs lost their thermal equilibrium at about 1.3 hours.

Things become even more interesting with the observational work which
have been done in the last decade. Verbunt (1997) re-analyzed the
period distribution of cataclysmic variables and came to the
conclusion that for Nova Like variables the period gap is not
significant. Different classes of variables were found to have
considerably different statistical properties. If the period gap does
not exist in the intrinsic period distribution of all variables, then
models which treat all CVs equally are potentially problematic.
Hellier \& Taylor (2000) re-examined the distribution of CVs. While
they argue against Verbunt's classification, their work confirms
Verbunt's conclusion that period gap is not significant for Nova-like
CVs.

Clemens et al. (1998) proposed an alternate model where the period gap
is produced by changes in the mass - radius relation that appear at
the edges of the gap.  However, this does not address the question of
why different types of CVs prefer one or another side of the gap.
Kolb, King \& Ritter (1998) demonstrated that a sudden change of the
mass-radius relationship of the Clemens et al. form would produce two
spikes rather than the observed distribution of CVs over period.  The
question of the impact of the mass-radius relationship on the
distribution of CVs is an interesting one that we will return to
later.

In this paper we concentrate on the question of the angular momentum
loss prescription which is appropriate for CVs.  The evolution of CVs
is closely related to the angular momentum loss,
and therefore it is important to have a correct prescription for
it. We introduce an empirical angular momentum loss rate
obtained from main sequence stars in open clusters with a range of
mass and age.  In section 2 we
describe the stellar model physics that we use, including a comparison
of the empirical angular momentum loss rate to the Rappaport, Verbunt,
\& Joss (1983) rates which are commonly used in the literature.  In section 3
we illustrate a method for inferring the time-averaged mass accretion
rate from knowledge of the angular momentum loss rate and the
mass-radius relationship.  In this section we also infer a mass-period
relationship and compare it to the observations.  We compare the
various timescales of interest for the evolution of CVs and conclude
that the secondary star in CV systems should be in thermal equilibrium
at and above the 2-3 hour period range.  Our results, along with a
discussion of the evolutionary state of the secondaries as a possible
explanation of the period gap,  can be found in section 4.

\section{Physics}

In this section we describe the physics of CVs relevant to our
calculations.\footnote{All capital letters in this section and farther
denote quantities in cgs units while all small letters express
quantities in dimensionless units relative to the sun: $
m=\frac{M}{M_{\odot}} , r=\frac{R}{R_{\odot}} $ } The properties of the
secondary star are important for understanding the time evolution in
CV systems.  We describe the global properties of our stellar models
in section 2.1.  We then briefly describe the angular momentum of the
binary system in section 2.2.  Section 2.3 is devoted to the important
issue of the angular momentum loss prescription for CVs.

\subsection{Stellar model}

We need the total moment of inertia and radius of the secondary stars
in CVs to determine the angular momentum evolution and the
mass-radius-period relationship for the system respectively.  In
addition, the luminosity is required to estimate the Kelvin-Helmholtz
timescale.

For the purposes of this paper we constructed a zero-age main sequence
set of models to determine the radius, luminosity, and moment of
inertia as a function of mass.  The angular momentum loss saturation
threshold $\omega_{crit}$ (see Section 2.3) as a function of mass was
taken from Sills, Pinsonneault \& Terndrup (2000).  The Yale Rotating
Stellar Evolution Code (YREC, Guenther et al. 1992) was used to
construct these models.  The nuclear reaction rates are taken from
Gruzinov \& Bahcall (1998). The heavy element mixture is that of
Grevesse \& Noels (1993), and our models have a metallicity of
Z=0.0188. Gravitational settling of helium and heavy elements is not
included in these models.

We use OPAL opacities (Iglesias \& Rogers 1996) for the interior of
the star down to temperatures of $\log T (K) = 4$. For lower
temperatures, we use the molecular opacities of Alexander \& Ferguson
(1994).  For regions of the star which are hotter than $\log T (K)
\geq 6$, we used the OPAL equation of state (Rogers, Swenson \&
Iglesias 1996). For regions where $\log T (K) \leq 5.5$, we used the
equation of state from Saumon, Chabrier \& Van Horn (1995), which
calculates particle densities for hydrogen and helium including
partial dissociation and ionization by both pressure and temperature.
In the transition region between these two temperatures, both
formulations are weighted with a ramp function and averaged.  The
equation of state includes both radiation pressure and electron
degeneracy pressure. For the surface boundary condition, we used the
stellar atmosphere models of Allard \& Hauschildt (1995), which
include molecular effects and are therefore relevant for low mass
stars.  We used the standard B\"{o}hm-Vitense mixing length theory
(Cox \& Guili 1968; B\"{o}hm-Vitense 1958) with $\alpha$=1.72.  This
value of $\alpha$, as well as the solar helium abundance,
$Y_{\odot}=0.273$, was obtained by calibrating models against
observations of the solar radius ($6.9598 \times 10^{10}$ cm) and
luminosity ($3.8515 \times 10^{33}$ erg/s) at the present age of the
Sun (4.57 Gyr).

The zero-age main sequence model properties define the normal single
star mass-moment of inertia, mass-radius, and mass-luminosity
relationships that we will apply to the study of CVs.

\centerline{
\begin{tabular}{|l|l|l|l|l|}\hline\hline
   m           & r         & l          & I[cgs]      & $\omega_{crit}    $
\\ \hline
   0.1         & 0.118     & 0.001      & 2.81e+51    & 9.50e-7             \\
   0.2         & 0.216     & 0.005      & 1.89e+52    & 2.31e-6             \\
   0.3         & 0.289     & 0.011      & 5.04e+52    & 3.69e-6             \\
   0.4         & 0.361     & 0.019      & 9.80e+52    & 5.13e-6             \\
   0.5         & 0.451     & 0.037      & 1.61e+53    & 6.76e-6             \\
   0.6         & 0.553     & 0.074      & 2.38e+53    & 1.16e-5             \\
   0.7         & 0.643     & 0.146      & 3.21e+53    & 1.45e-5             \\
   0.8         & 0.711     & 0.266      & 4.16e+53    & 1.82e-5             \\
   0.9         & 0.785     & 0.444      & 5.36e+53    & 2.27e-5             \\
   1.0         & 0.883     & 0.690      & 6.85e+53    & 3.00e-5             \\
   1.1         & 0.995     & 1.028      & 8.61e+53    & 4.39e-5             \\
   1.2         & 1.119     & 1.476      & 1.06e+54    & 8.66e-5             \\
\hline\hline
\end{tabular}
}

Table 1 gives the radius and luminosity in solar units and the cgs moment
of inertia as a function of the mass in solar masses.  The important
saturation threshold $\omega_{crit}$ is discussed in section 2.3.
We interpolate in the above table to get stellar properties as a
function of mass.  Some investigators use linear or power-law fits for
these global properties.  The mass-radius relationship for unevolved
secondary stars can be expressed as $r\approx 0.901m+0.014$.  The
mass-luminosity relationship can be fit by a broken power law of the
form $ \lg{(l)}\approx 2.022\lg{(m)}-0.878 , m\leq 0.5 ;
4.730\lg{(m)}-0.117 , m\geq 0.5 $.  Finally, the moment of inertia as
function of mass is described by $\lg{I} \approx 2.397 \lg{m}+ 53.91$.

\subsection{Angular momentum}

The timescale for tidal synchronization is much shorter than the
characteristic evolutionary timescale for the orbital period range of
cataclysmic variables (see section 3.)  We can therefore assume that
the rotation periods of the stars and the system are the same; because
of the small moment of inertia of the white dwarf its rotational
angular momentum can be safely neglected.  Helioseismic data for the
Sun indicates that the rotation rate in the surface convection zone is
independent of radius (e.g. Schou et al. 1998)
Most helioseismic inversions indicate that the rotation in the solar
core is similar to that of the surface convection zone down to a depth
of order 0.2 solar radii (Chaplin et al 1999);
there is some debate about the situation in deeper layers (compare
Chaplin et al. 1999 with  Gavryuseva, Gavryusev, \& di Mauro,
2000)   The stars in CV systems rotate
relatively rapidly (implying a short timescale for internal angular
momentum transport) and have deep surface convection zones, so it is
therefore reasonable to assume solid-body rotation in the secondary
star.

The total angular momentum of the system can thus be written in the form :
\begin{equation}
J=M_{\odot}^{5/3}G^{2/3}m_1 m_2 m^{-1/3}\omega^{-1/3}+I(m_2)\omega
\end{equation}

where $m_1$ and $m_2$ are the masses of the primary and secondary, $m$
is the total mass of the system, $\omega$ is the angular rotation
velocity of the secondary, and $I(m_2)$ is the moment of inertia of
the secondary star. The first term describes the orbital angular
momentum of the system, and the second term is the spin angular
momentum of the secondary star.

\subsection{Angular momentum loss}

The evolution of a CV is determined by the masses of the
primary and secondary, the structural properties of the secondary star, and the
angular momentum loss rate. We will show in section 3.5
that the mass loss rate can be inferred from the mass-radius
relationship and knowledge of the angular momentum loss rate under the
assumption of marginal contact.  The angular momentum loss
prescription is therefore a crucial ingredient for the study of CVs.

There are two general mechanisms for the transferring angular momentum
out of the system. 

First, binary systems emit gravitational radiation, which carries away
angular momentum.  The rate of angular momentum loss increases as the
orbital separation decreases, but it decreases as the total mass of
the system decreases.  Because both the secondary mass and the orbital
separation decrease as CV systems evolve, the net effect is a loss
mechanism that does not depend strongly on the orbital period.

We use the following expression for the angular momentum loss from
gravitational radiation (see, for example, Landau \& Lifshitz 1939):

\begin{equation}
\left(
\frac{dJ}{dt}
\right)_{grav}
=-\frac{32}{5} \frac{G^{7/2}}{c^5} a^{-7/2} m_1^2 m_2^2 \sqrt{m}
M_{\odot}^{5/2}
\end{equation}
Where $m_1$ , $m_2$ , $m$ are the white dwarf mass, secondary mass,
and total mass respectively; $a$ is the separation between the stars,
which can be obtained from Newtons' form of Kepler's third law
$a=\left( \frac{G m M_{\odot}}{\omega^2} \right)^{\frac{1}{3}} $.

However, gravitational radiation alone cannot explain the orbital
evolution of low mass stars (e.g. Patterson 1984, Rappaport, Verbunt
\& Joss 1983.)  Secondary stars with sufficiently deep surface
convection zones experience angular momentum loss from a magnetic
stellar wind.  Because binaries at the orbital periods of CVs are
tidally synchronized (Patterson 1984), angular momentum lost from the
secondary star is removed from the orbital angular momentum and the
orbital separation of the binary system is reduced.  Weber \& Davis
(1967) predicted an angular momentum loss rate proportional to
$\omega^3$ based upon a study of the solar wind; this is consistent
with the time dependence of rotation inferred from early studies of
solar-type stars in open clusters (Skumanich 1972).  The strong
dependence of the angular momentum loss rate on the angular velocity
results primarily from scaling the mean magnetic field strength to the
rotation rate.  Rappaport, Verbunt \&Joss (1983) developed an
empirical prescription that is commonly used in studies of CVs; their
relationship is given by:

\begin{equation}
\left(\frac{dJ}{dt}\right)_{wind}\approx -3.8\cdot 10^{-30} \cdot 
M_{\odot}R_{\odot}^4\cdot m
r^{\gamma}\omega^3\,dyn\,cm.
\end{equation}
where $\gamma$ is a dimensionless parameter in the range from 0 to 4.

There are serious difficulties with applying any angular momentum loss
model which scales as $\omega^3$ to observations of young low mass
stars.  The spindown of rapid rotators is predicted to be extremely
fast (see for example Pinsonneault, Kawaler, \& Demarque 1990), but
rapidly rotating low mass stars are observed in young open clusters
(Stauffer \& Hartmann 1988).  Both chromospheric activity indicators
and X-ray studies, furthermore, can be used as proxies for
measurements of the strength of stellar magnetic fields.  There is now
extensive empirical evidence that both chromospheric and coronal
indicators become independent of the rotation rate above a
mass-dependent critical angular velocity (e.g. Patten \& Simon 1996);
this would imply a much lower angular momentum loss rate for rapid
rotators.  We note that this does not contradict the overall Weber \&
Davis (1967) model in the case of slow rotators; however all of the
direct and indirect observational tests in low mass stars and open
clusters of different ages indicates that it overestimates the angular
momentum loss rate for rotation periods shorter than 2.5 - 5 days.

A number of different theoretical groups have investigated the
spindown of low mass stars (Queloz et al. 1998; Collier-Cameron \&
Jianke 1994; Keppens, MacGregor, \& Charbonneau, 1995; Krishnamurthi
et al. 1997; Sills, Pinsonneault, \&
Terndrup, 2000).  In all cases the survival of rapid rotation in young
open clusters required a modification of the angular momentum loss law
at high rotation rates.  We therefore use an angular momentum loss
prescription with the same function form as that of Sills, Pinsonneault,
\& Terndrup (2000):

\begin{equation}
\left(\frac{dJ}{dt}\right)_{wind} = -K_w \cdot \sqrt{\frac{r}{m}}\cdot
\left\{
\begin{array}{ll}
\omega^3                & {\rm for} \; \omega \leq \omega_{crit} \\
\omega \omega^2_{crit}  & {\rm for} \; \omega > \omega_{crit}
\end{array}
\right.
\end{equation}

Here $\omega_{crit}$ is the critical angular frequency at which the
angular momentum loss rate enters into the saturated regime.  The
constant $K_w\approx2.7\cdot10^{47}\,g\,cm\,s$ and is calibrated to
reproduce the known solar rotation period at the age of the Sun (see
Kawaler 1988 for a discussion of the ingredients and uncertainties.)
The value of $\omega_{crit}$ can be inferred empirically by
reproducing the observed spindown of low mass stars as a function of
mass and age.  There is therefore one clear implication from the large
database of observational and theoretical work in the study of
rotation in low mass open cluster stars: angular momentum loss
prescriptions of the form used by Rappaport, Verbunt \& Joss (1983)
greatly overestimate angular momentum loss rates for secondary stars
in the period range of interest for the study of CVs.
Solokani, Motamen \& Keppens (1997) propose an alternate
physical mechanism (concentration of magnetic fields close to the pole
for rapid rotators), but their model predicts a very similar angular
momentum loss rate to the saturated wind model.

The second important result in the context of CV studies is the
observed mass dependence of the angular momentum loss rate.  It has
long been known that the observed saturation threshold for
chromospheric and coronal activity indicators is mass dependent (see
Noyes et al. 1984; Patten \& Simon 1996).  Krishnamurthi et
al.  (1997) found that a scaling of $\omega_{crit}$ with the inverse
Rossby number (the ratio of the rotation period to the convective
overturn timescale) gave a reasonable fit to the observed timescale
for spindown as a function of mass from 0.6 - 1.2 solar masses.
Similar Rossby number scalings are also found to describe the
saturation of activity indicators (Krishnamurthi et al. 1998) and are
expected on general theoretical grounds for a shell dynamo
(e.g. Durney \& Latour 1978). In other words, the timescale
for stars to spin down increases as the mass decreases.

Observations of both activity indicators and rotation velocities down
to very low masses have been obtained in open clusters with a range of
ages (Jones, Fischer, \& Stauffer 1996; Stauffer et
al. 1997; Terndrup et al. 2000; Reid \& Mahoney 2000).
There is no observational support for an abrupt change in angular
momentum loss properties at the point where stars become fully
convective (around 0.3 solar masses.)  Sills, Pinsonneault, \&
Terndrup (2000) found that the mass dependence of $\omega_{crit}$
could no longer be fit by a Rossby scaling below 0.5 solar masses;
this suggests that the transition from a shell to a distributed dynamo
occurs well above the fully convective boundary.  However, some
angular momentum loss was needed even below the fully convective
boundary in order to explain the observed spindown of Hyades stars
relative to Pleiades stars.  Hawley (1999) and Hawley et al. (1999) 
also found that the timescale for high activity levels to
survive was a smooth function of mass.

\begin{figure}[t]
\plotone{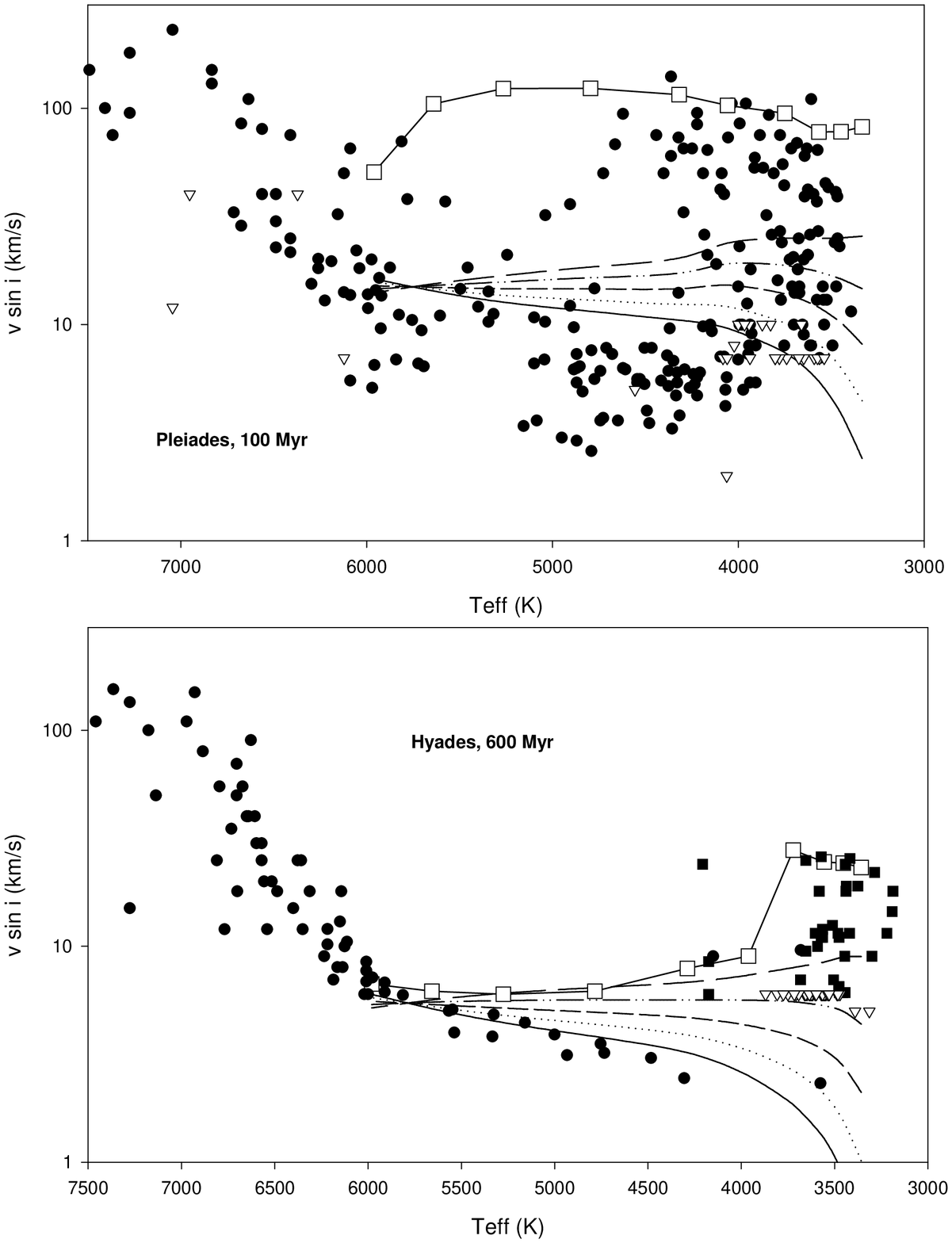}
\caption{\tiny Theoretical models for the upper envelope of rotation compared with data in the Pleiades (top)
and Hyades (bottom). The solid line with the open boxes is the empirical model (0.2 - 1.1 $M_\odot$ in the 0.1 $M_\odot$
increments). The group of lines below are the RVJ models with $\gamma =0,1,2,3,4$. Pleiades data is taken from Terndrup et al.
 2000, Stauffer et al. 1997, and Anderson et al. 1966. Hyades data is
taken from Radick et al. 1987, Terndrup et al. 2001, and Kraft 1965.}
\end{figure}

We compare the predicted stellar rotation rates using the Sills,
Pinsonneault, \& Terndrup (2000) empirical angular momentum loss law
as a function of mass and the Rappaport, Verbunt, \& Joss (1983)
angular momentum loss law with Pleiades and Hyades data in Figure 2.
For these models we chose an initial rotation period at the
deuterium-burning birthline of 10 days (corresponding to the average
rotation period inferred for T Tauri stars from Choi \& Herbst 1996).
Because of the strong feedback in the unsaturated angular momentum
loss law, choosing a much shorter initial rotation period would yield
very similar results for the Rappaport et al. loss rates.  This choice
of initial conditions corresponds to the expected upper
envelope of rotation rates that are observed in open clusters.

We note that because the most rapid rotators are rare (roughly 3\% of
the total population) the upper envelope as a function of effective 
temperature is subject to Poisson noise.  However, the existence of
stars rotating more rapidly than 10 km/s in the Pleiades is a direct 
contradiction of the predictions of an unsaturated angular momentum
loss law.  We do see evidence of a transition away from a pure Rossby 
scaling of the saturation threshold in the coolest Hyades stars.  The
upturn in rotation at the low mass end indicates a change in the 
efficiency of angular momentum loss (a pure Rossby scaling would
reduce to the older angular momentum loss rates at this age.)
However, this transition occurs at 0.6 solar masses, well above the
fully convective boundary.  Furthermore, there is clear evidence for 
spindown even in the lowest mass stars for which we have data.  This 
extends into the very low mass regime from field star data (Hawley et 
al. 1999.)  

We compare our empirical angular momentum loss rates as a function of
secondary mass with that from the Rappaport et al.(1983) prescription in
Figure 3; in the latter prescription magnetic braking is stopped at
0.3 solar masses and the only angular momentum loss mechanism for the
low mass stars is gravitational radiation.
The older angular momentum loss rate is overestimated by
about 2 orders of magnitude at the period gap.

\begin{figure}[t]
\plotone{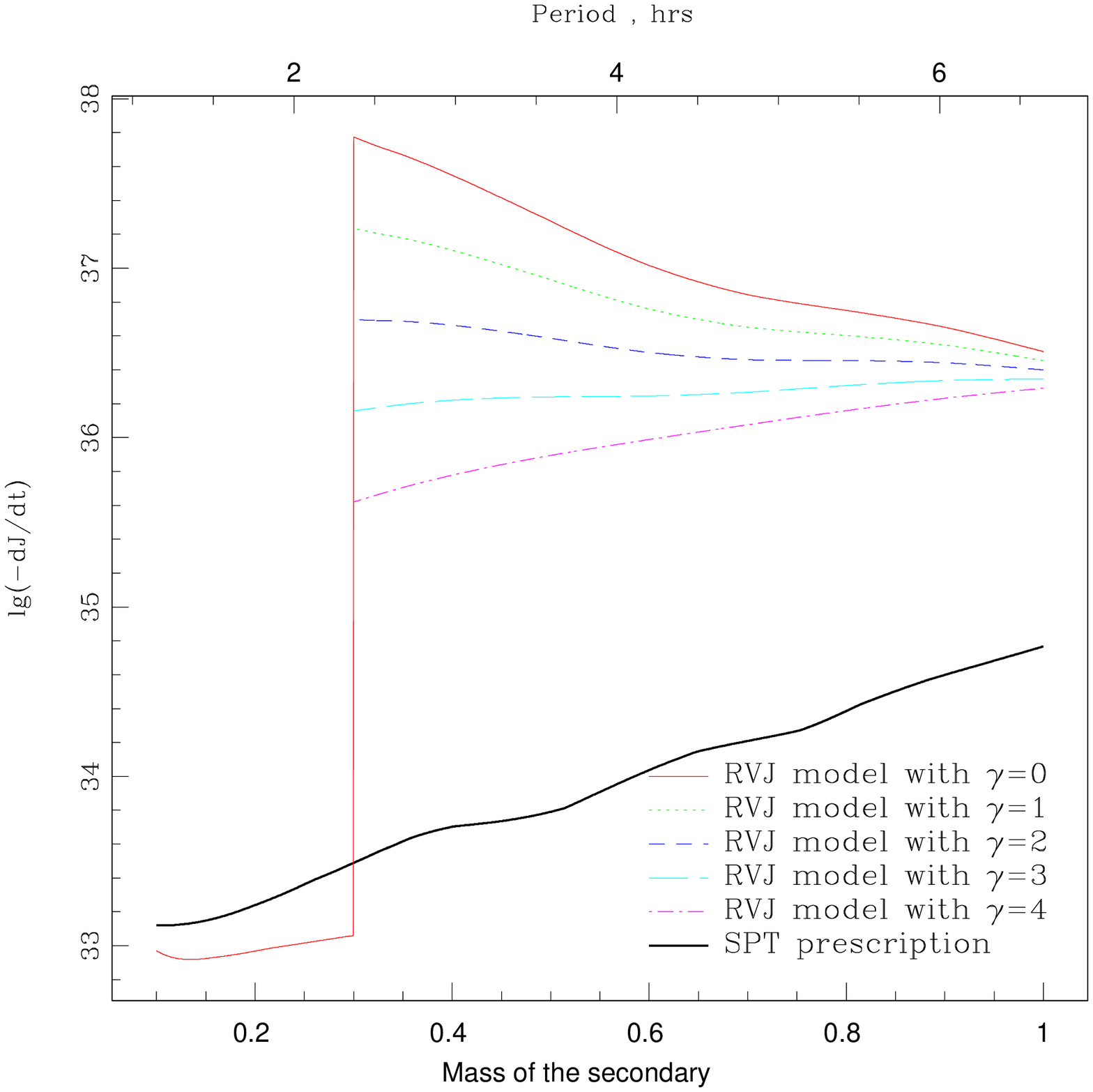}
\caption{\tiny Angular momentum loss for a system with mass-conserving
primary of 0.62$M_\odot$. Dashed lines are Rappaport,Verbunt\&Joss
model, thick solid line is Sills, Pinsonneault \& Terndrup (2000)
prescription.}
\end{figure}

We note that the stellar data becomes sparse below about 0.2 solar
masses, and further observational data on very low mass stars would be
useful to constrain the empirical angular momentum loss law at the
bottom of the main sequence.  It is also possible that for the
faintest stars the timescale for angular momentum loss could become
much longer (see Basri \& Marcy 1995 for a discussion of this point.)
Any such effect, however, occurs at a significantly lower mass than
the fully convective boundary and therefore has no direct effect on
the existence of the period gap.  This change in the angular momentum
loss law has important consequences for the predicted time evolution
of CVs which we present in the next section.

\section{Model, limitations and solution.}
This section describes our model for the evolution of CVs.  Section
3.1 describes our mass-radius relation, which is one of the most important
ingredients in the evolution of cataclysmic variables.  We summarize
all of the equations and assumptions in our models in Section 3.2.
Section 3.3 gives a motivation for the assumption of marginal
contact.  We compares our inferred theoretical mass-period relation to
the empirical one in section 3.4, and discuss the time-averaged mass
accretion rate as a function of period in section 3.5.  Section 3.6 is
devoted to a comparison of the timescales of different processes.

\subsection{Mass - radius relation for the secondary}

In this work we begin with the theoretical mass-radius relationship
for single unevolved low mass main sequence stars described in section
2.  These models provide a natural starting point for the study of
CVs; if the secondary star begins with a mass at or lower than
$\sim$0.8 solar masses then there will be little nuclear evolution
within a Hubble time and such objects would therefore be expected to
follow the unevolved mass-radius relationship.  In principle, both the
rapid rotation of CV secondaries and the tidal gravitational field of
the white dwarf primary could affect the mass-radius relationship.
However, Kolb \& Baraffe (1999) investigated the impact of rotation
and tidal distortion on the properties of CV secondaries. They
concluded that these effects are small even in short period CVs. We
therefore do not include the structural effects of rotation and tidal
deformation in our models. We do note, however, that if the secondary
star is not chemically homogeneous then meridional circulation and
rotational instabilities could induce significant mixing; as we will
see below, there is possible evidence that some CV secondaries could
have experienced significant nuclear evolution on the main sequence.
We defer a treatment of these effects to a paper in preparation.

The secondary star in CVs could also be partially evolved.  This could
occur if the secondary and primary stars had similar mass;
alternately, if the timescale for CV evolution was sufficiently long,
primary stars more massive than $\sim 1$ solar mass could experience
significant nuclear burning during the CV phase.  A proper treatment
of this effect would require calculation of a range of initial secondary
masses and evolutionary states and stellar interiors models which
evolve under the presence of mass loss; we are constructing such
models in a paper in preparation.  We have also considered two
simplified models which do not begin in a chemically homogeneous state
to test for the impact of the evolutionary state of the primary on the
mass-radius relationship.  We evolved 1.0 and 1.2 solar mass, solar
composition models until they overflowed their Roche lobes at an
initial orbital period of 10 hours for an assumed primary white dwarf
mass of 0.6 solar masses.  We then removed mass from them
without further nuclear burning; these models are systematically
larger at a given mass than the chemically unevolved models until they
become fully mixed at low total stellar mass.

\subsection{Model and limitations}
Our system of equations is:
\begin{equation}
\left\{
\begin{array}{l}
J=M_{\odot}^{5/3}G^{2/3}m_1 m_2 m^{-1/3}\omega^{-1/3}+I(m_2)\omega \\
\frac{dJ}{dt} = \left(\frac{dJ}{dt}\right)_w+\left(\frac{dJ}{dt}\right)_g \\
r_2=f(m_2)
\end{array}
\right.
\end{equation}

When the mass of the secondary drops below $\sim 0.07 M_\odot$ the
radius increases with decreased mass (see for instance Burrows et
al. 1993).  This would result in increased period for decreased mass;
systems which pass this critical point are sometimes called period
bouncers in the literature.  We do not consider secondaries with
masses below this limit.

Accretion happens only when the secondary star overfills its Roche lobe. In
order to take
this into account we introduce a dimensionless
parameter $y$ which is defined as the ratio of the
star's radius to the Roche distance from the center of the secondary.
For the Roche distance from the secondary we use the relation derived
by Eggleton (1983):
\begin{equation}
y=\frac{R}{R_L}=\frac{R_\odot r_2}{a}\cdot
\frac{0.6\left(\frac{m_2}{m_1}\right)^{2/3}+\ln{\left[1+\left(
\frac{m_2}{m_1} \right)^{1/3}\right]}}
{0.49 \left(\frac{m_2}{m_1}\right)^{2/3}}
\end{equation}

Accretion only occurs if $y\geq 1$.  In the
marginal contact assumption described below, we constrain $y$ to be unity.

\subsection{Marginal contact}

The usual way to make the system of equations (10) closed is to add a
prescription for the mass accretion rate
(D'Antona et al 1989, Hameury 1991). This is obtained indirectly
from observations (see Patterson 1984, Rutten, Paradijs and Tinbergen
1992), or from semi-analytic expressions (D'Antona et al 1989).
The uncertainties in the observed mass loss rates are significant; the
estimates of mass accretion are based on
poorly understood physics of accretion disks
and the distances to CVs are not well constrained (Patterson 1984,
Rutten et al 1991).

Directly using the inferred mass loss rates
leads to oscillations of the period in a small timescale (Hameury 1991).
Due to orbital angular momentum loss, the period of the system decays until
the secondary overfills its Roche lobe and mass loss sets in. When
mass is lost from the system the moment of inertia is reduced, which
causes the orbital
period to increase. If the timescale for mass accretion is short
enough this can drive the system out of contact and lead to
oscillations in the orbital period.  To resolve this instability some
investigators have introduced a
dependence of the mass accretion rate on the extent to which the Roche
lobe is overfilled (Hameury 1991, D'Antona et al. 1989).

However, we contend that knowledge of the angular momentum loss rate
and the mass-radius relationship specifies the time-averaged mass loss
rate with a small uncertainty.  CV systems will tend
to remain close to marginal contact; a small decrease in orbital
period will drastically increase the mass loss rate and drive the
stars apart because of the decrease in moment of inertia, while an
increase in orbital period will drive $y$ below
unity and detach the stars, causing the stars to come closer from
angular momentum loss.  For any given primary and secondary masses,
this implies that the orbital period should be close to the case where
$y$ is unity.  If the mass-radius relationship is known, this can be
directly converted into a mass-orbital period relationship.
There is also a similar and well constrained relationship
between the total angular momentum and the orbital period.

There is no strong motivation for the existence of changes of mass
accretion in a small timescale. But even if such oscillations exist,
they would not make difference in time averaged angular momentum loss
rate. First, such oscillations would be of order of the timescale for
the accretion disk, which is much lower than any other characteristic
time for CVs. Secondly, the weaker dependence of the saturated angular
momentum loss rate upon $\omega$ than in the older $\dot{J}$
prescriptions would imply a smaller amplitude of fluctuations of
angular momentum loss. Therefore, when averaged over a time larger
than the characteristic time for the accretion disk, the angular
momentum rate would not be very sensitive to such oscillations.

We therefore introduce the assumption of marginal contact; namely that
the secondary is always at the critical orbital period where it begins
to overflow its Roche lobe ($y=1$).  The time-averaged mass accretion
rate is inferred from the angular momentum loss rate and the
mass-radius relationship.  This can be regarded as a steady-state
accretion rate.  It is mathematically stable; of course, this does not
guarantee the physical stability of such a flow in a real accretion
disk.  For understanding the period distribution of CVs, however, this
should be a good approximation.  A comparison of the steady-state mass
loss rate to the observed one, and its implications, are discussed in
section 3.6.

The mass-radius relationship that we use is for secondary stars in
thermal equilibrium; this assumption can be tested in two different
ways.  In the next section we compare our mass-period relationship for
both unevolved and evolved secondary stars to the observed one; in
section 3.5 we compare the Kelvin-Helmholtz timescale to the
timescales for angular momentum and mass loss.

\subsection{The Secondary Mass-Orbital Period Relationship}

The mass-period relation for the secondary can be obtained
in the assumption of marginal contact with one final ingredient: the
amount of mass from the secondary that the primary retains.
The rates of change of the primary and secondary masses are related by

\begin{equation}
\frac{dm_1}{dt}=-\chi\frac{dm_2}{dt}
\end{equation}
where $\chi$ is a dimensionless parameter that varies from 0 to 1.
In the case of $\chi=1$ the system conserves its total mass.  In the
case of $\chi=0$ the primary mass is constant and the rest of the mass
is lost in nova outbursts.

If one specifies the masses and the value of $\chi$, for a given
mass-radius relation and angular momentum loss rate, the period is
defined.  This yields a relationship between the mass of the secondary
and period for CVs.  The rate of period change is correlated with the
mass accretion rate through the mass of the secondary-period relation:

\begin{equation}
\frac{d\omega}{dt}=\left.
\dot{m_2}\frac{d\omega}{dm_2}\right|_{m_2+\chi m_1=const}
\end{equation}

By requiring $y=1$ we obtain the relation:
\begin{equation}
R_\odot r_2(m_2) = \left(
\frac{G (m_1+m_2 )M_{\odot}}{\omega^2}
\right)^{\frac{1}{3}}
\cdot \frac{0.49
\left(\frac{m_2}{m_1}\right)^{2/3}}{0.6\left(\frac{m_2}{m_1}\right)^{2/3}+
\ln{\left[1+\left( \frac{m_2}{m_1} \right)^{1/3}\right]}}
\end{equation}

The mass-period relation is shown in figure 4 together with two empirical fits
$
m_2=(0.038\pm 0.003)P^{1.58\pm0.09};
m_2=(0.126\pm 0.011)P-(0.11\pm0.04)
$
taken from recent work on CV secondaries by Smith and Dhillon (1999).

\begin{figure}[t]
\plotone{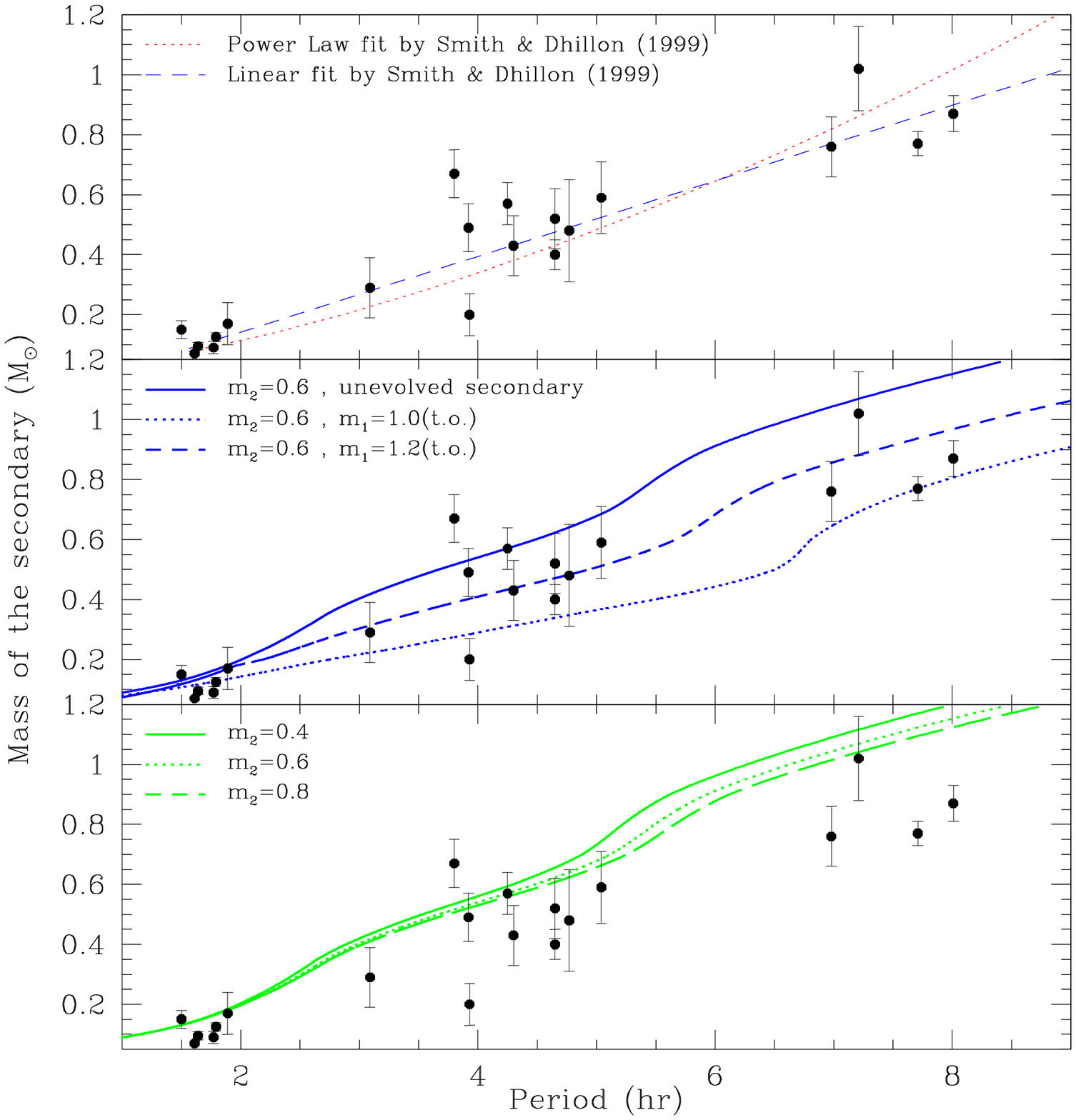}
\caption{\tiny The mass of the secondary - period relation.}
\end{figure}

In the top panel we show the data and empirical fits.  The bottom
panel illustrates the effect of changing the primary mass in the range
of 0.4 to 0.8 solar masses; this has only a small impact on the
predicted mass-period relationship.  In the middle panel we compare
our unevolved secondary models and two simplified evolved secondary
models with the data.

The unevolved relationship is a good match to the upper envelope of
the data.  Especially at longer periods, there is a tendency for some
of the data to lie at lower mass for a fixed period, which indicates a
larger radius than predicted by the unevolved models.
However, the observed range of mass-period relationships is consistent
with the range spanned by our partially evolved and unevolved models.

At short periods, the M-R relationship for our evolved models could be
altered because there is significant helium enrichment and we are
using model atmospheres with a solar mix. Even though the predictions
are close to the data around periods of two hours, the error bars
there are smaller. This could also indicate that non-spherical effects
become important at low period, which is plausible as those stars
could be severely distorted by their much more massive and close WD
companion.

In conclusion, the models are consistent with the data if a range of
secondary evolutionary states are considered.  There is evidence for a
departure from thermal balance only if the unevolved models are
considered as the only ones.  We directly compare the relevant
timescales in the next section.  We also compare the data with the
inferred mass loss rates for different mass-radius relationships in
the section that follows.  We contend that these diagnostics are more
consistent with a range of evolutionary states rather than a departure
from thermal balance.

\subsection{Timescales}

The governing timescale for the orbital evolution of CVs is the
timescale for angular momentum loss.  Our mass-radius relationship
assumes models in thermal equilibrium, so it is important to compare
the timescales for orbital change and for secondaries to establish
thermal balance (e.g. the Kelvin-Helmholtz timescale.)  In addition,
we can infer the steady-state accretion rate and the timescale for
mass loss from the secondary.

For angular momentum loss, mass accretion and period change timescales
can be estimated by $ \tau_X = \left| \frac{X}{\dot{X}}\right| $
where $\dot{X}$ is the time derivative of variable $X$.

The steady-state mass loss rate can be derived as follows.
Taking the derivative from the angular momentum equation (1) and
using equation (8), one can get :
\begin{equation}
\begin{array}{l}
\dot{J}=\dot{m_2}\left( A\omega^{-1/3}m^{-1/3}(m_1-\chi m_2)-
\frac{1}{3}Am_1 m_2 m^{-4/3}\omega^{-1/3}(1-\chi)\right.\\
-\frac{1}{3} A m_1 m_2 m^{-1/3}\omega^{-4/3}\frac{d\omega}{dm_2}+
\left. I(m_2)\frac{d\omega}{dm_2}+\omega\frac{dI}{dm_2}\right)
\end{array}
\end{equation}

We can now solve for $\dot{m_2}$ because we know the angular momentum
loss rate, moment of inertia as a function of mass and the mass-period
relationship, which gives $\frac{d\omega}{dm_2}$:
\begin{equation}
\begin{array}{l}
\dot{m_2}=\dot{J}\left(A\omega^{-1/3}m^{-1/3}(m_1-\chi m_2)-\frac{1}{3}Am_1
m_2 m^{-4/3}
\omega^{-1/3}(1-\chi)\right.\\
-\frac{1}{3}A m_1 m_2 m^{-1/3}\omega^{-4/3}\frac{d\omega}{dm_2}
\left.+I(m_2)\frac{d\omega}{dm_2}+\omega\frac{dI}{dm_2}\right)^{-1}
\end{array}
\end{equation}
This defines the steady-state mass accretion rate directly, and the
rate of change of angular frequency through equation (7).
We solve this set of equations iteratively.


For CVs the timescale for tidal synchronization is very short, of
order $5000$ years (Warner 1995); this timescale is much shorter than
the other relevant ones for the system. The thermal relaxation or
Kelvin - Helmholtz timescale for the secondary is
$$
\tau_{KH}\approx \frac{3}{4}\frac{GM^2}{LR}
$$

We use the mass-radius and mass-luminosity relationships described
earlier. The relevant timescales for cataclysmic
variables are illustrated in figure 5.

\begin{figure}[t]
\plotone{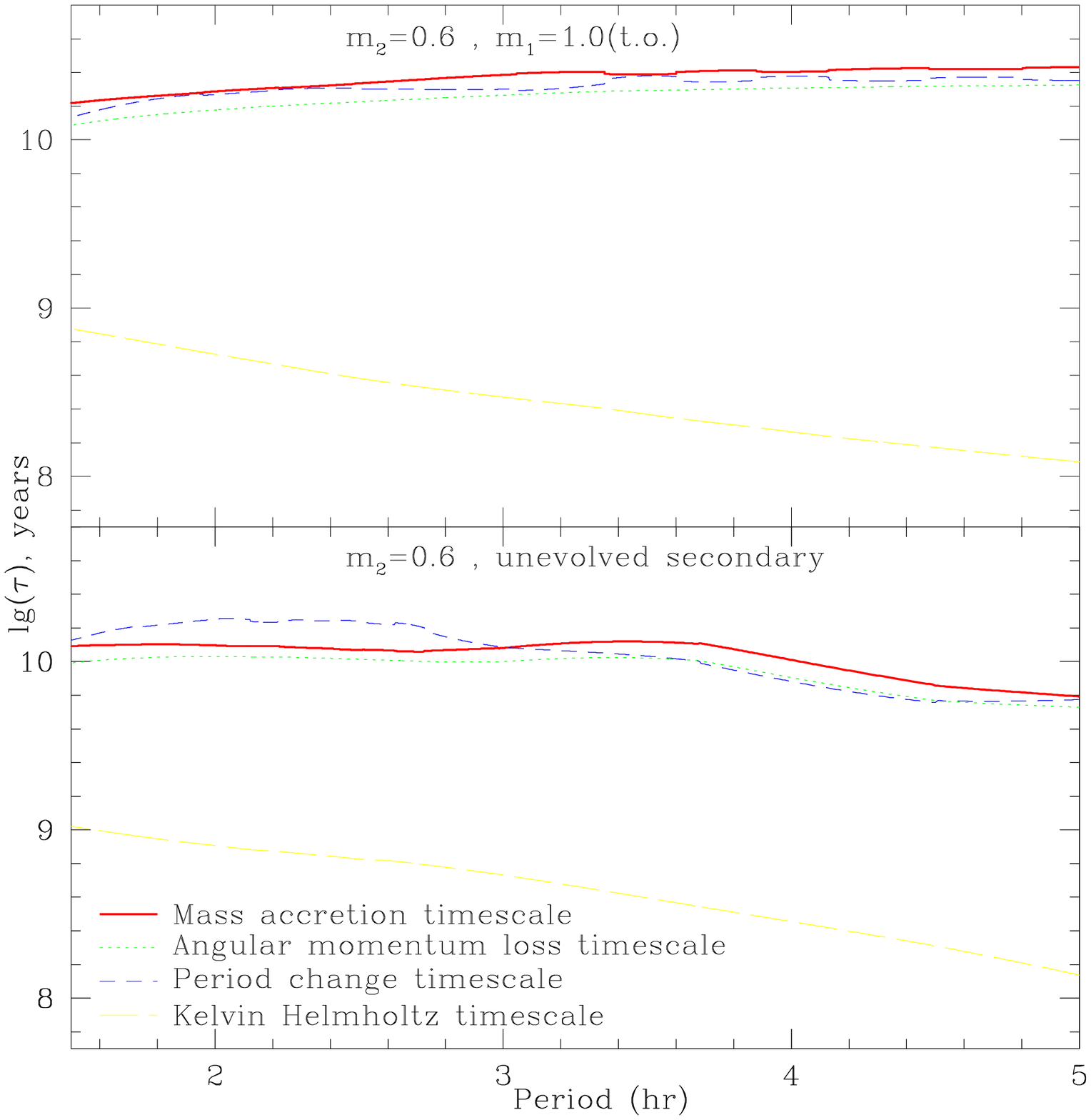}
\caption{\tiny Comparison of timescales.}
\end{figure}

The timescales for mass accretion and period change are comparable and
follow the timescale for angular momentum loss, because in the
assumption of marginal contact they are both defined by it. The
Kelvin-Helmholtz timescale is shorter; therefore we can infer that
the secondary stars in CVs should be in thermal equilibrium at all periods.
We also note that the timescales for the evolution of CVs are much
longer than in previous studies, which will have a profound impact. It
is no longer correct to assume that the timescale for CVs to reach the
period gap is less than a Hubble time.  Rather than treating the
period distribution of CVs as being in a steady-state, this implies
that finite age effects could be crucial.

Our timescale estimates imply that the secondary stars are in thermal
equilibrium which is consistent with unperturbed mass-radius
relationships that we have used. This is in contrast with the results
from earlier loss laws, and it is a straightforward consequence of our
angular momentum loss prescription. Early prescriptions for angular
momentum loss predicted much higher loss rates for rapid rotators than
are consistent with modern open cluster data.  A relatively low
angular momentum loss of the form we advocate leads to a low mass
accretion rate and a longer evolutionary timescale.  In addition, the
open cluster data indicate that angular momentum loss works even for
stars below the fully convective boundary.

This supports the idea that a range of evolutionary states, rather
than a departure from thermal balance, is responsible for the stars
with low mass at fixed period discussed in the previous section.
In a recent paper Rubenstein(2001) demonstrated that evolutionary stage and 
metallicity are able
to explain the observed Period-Color relation for contact binaries.  Our 
results are consistent with this work, although
we are describing the mass - period relationship. The closer correspondence 
between the observed properties of CVs and evolved stellar models has been 
noted earlier (e.g. Patterson 1984; Baraffe \&  Kolb 2000).

\subsection{Comparison of Mass Accretion Rates}

In this section we compare our mass accretion rates to the data.
We begin by noting that the uncertainties in the empirically derived mass 
accretion rates are significant (Patterson 1984).  First, it can be 
difficult to infer the luminosity.  The distances to CVs are not well 
constrained.  The bolometric correction is based upon the properties of 
x-ray emitting systems and involves the complex physics of accretion 
disks.  Knowledge of the luminosities of CV systems is therefore non-trivial.

Accretion rates are derived from the luminosity via the expression
$L_{bol}=kGM_1\dot{M}/R_*$ (Patterson 1984) which involves an
additional unknown: the parameter $k$ is efficiency of conversion of
gravitational energy to radiation. Patterson is using the low spatial
abundance of CVs as one of the arguments justifying high mass
accretion rate.  With a high mass accretion rate, CV systems will die
rapidly. However, we suggest that the solution for the slow spatial
abundance of CVs is actually slow angular momentum evolution. Systems
with relatively high orbital periods after the common envelope phase
($\sim$3-5 days) would not come into contact and hence most of them
exist as planetary nebulae with a companion, rather than CVs.

\begin{figure}[t]
\plottwo {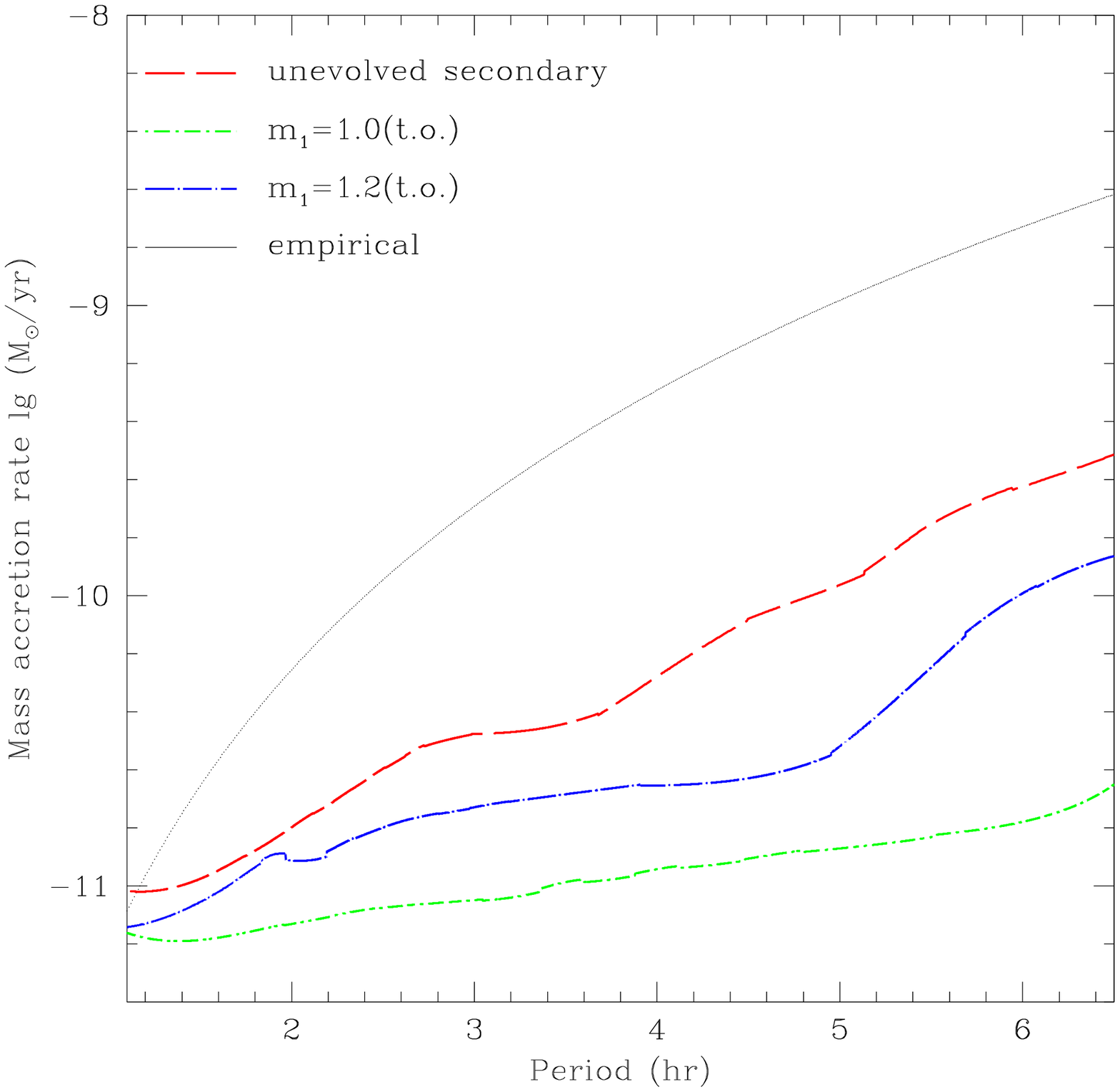}{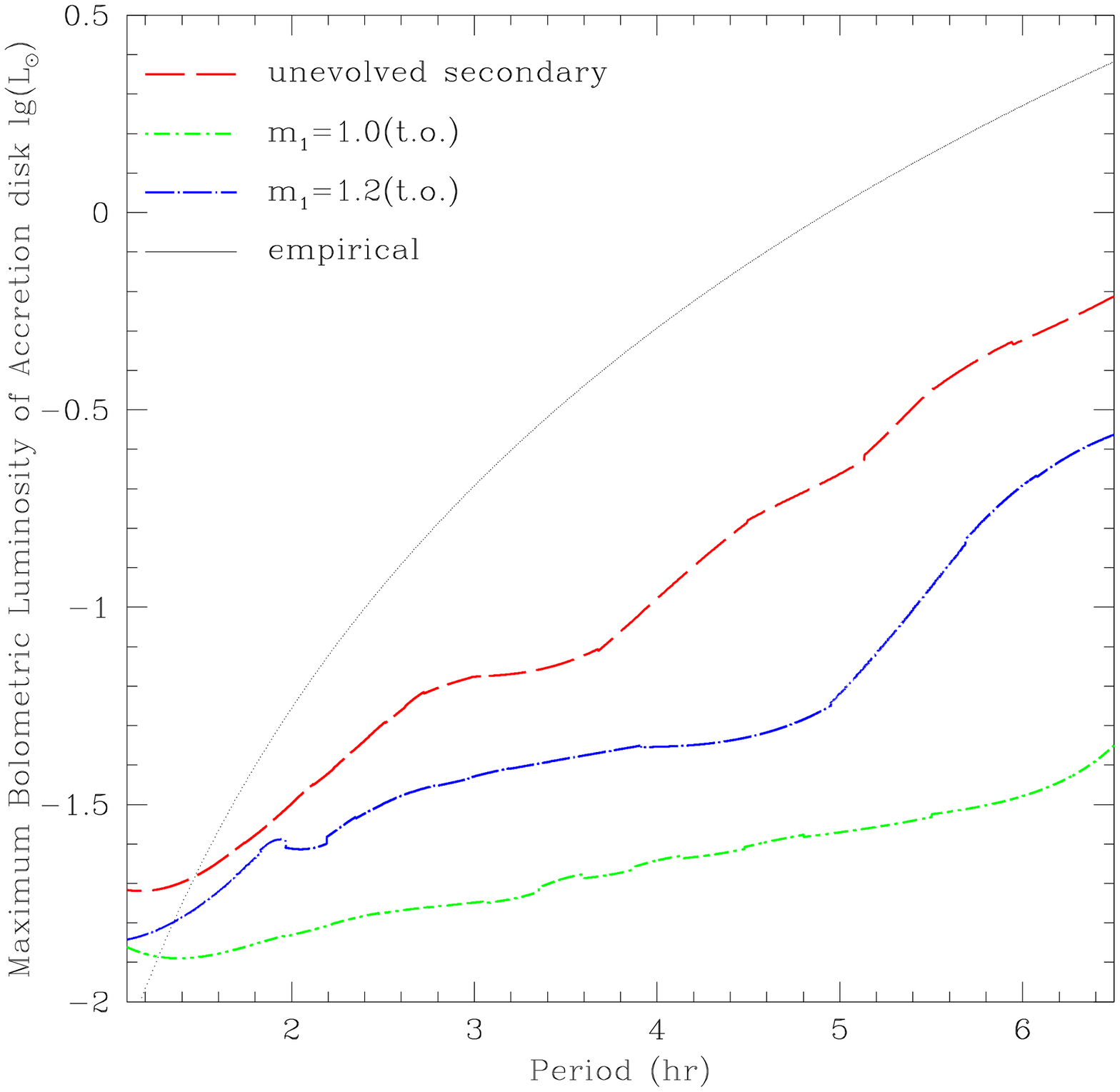}
\caption{\tiny Comparison of the time averaged mass accretion rate and maximum 
possible accretion disk luminosity for our 3 models.
The thin solid line is the empirical fit from Patterson (1984).}
\end{figure}

The theoretical mass accretion rate is derived using formula (10) assuming 
the maximum possible theoretical luminosity of the accretion
disk ($k=1$).  We compare the mass loss rates inferred from both the 
unevolved and evolved M-R relationships to the empirical results by 
Patterson (1984) in figure 6.
The inferred observed rates are closer to the evolved than unevolved 
models, but the empirical results are systematically significantly higher 
than the theoretical predictions in both cases.  There are three possible 
classes of explanations.  First, this could reflect systematic errors in 
the measured rates.  Alternately, this could indicate a problem with the 
mass-radius relationship; as in the case of the mass-period relationship we 
find that evolved secondaries provide a better fit to the data.  Finally, 
there is an difference between the {\it instantaneous} mass loss rate that 
is observed and the {\it time-averaged} mass loss rate obtained under 
marginal contact.  This discrepancy could therefore reflect a duty cycle, 
where the systems are only in contact (and visible as CVs) for a fraction 
of their total lifetime.
With better distance estimates and new x-ray satellites
it will be possible to distinguish between these different explanations.

\section{Summary}

The most important result of this paper comes from applying the
empirical angular momentum loss rates inferred from young single stars
to the evolution of cataclysmic variables.  In contrast to some
popular models, we find that the timescale for angular momentum loss
from CV secondaries is significantly longer than previously assumed.
This has two direct and important consequences.

First, the angular momentum loss timescale is significantly longer than the
Kelvin-Helmholtz timescale for all but the shortest period CVs, which
implies that these objects are in thermal balance and not dramatically
increased in radius relative to normal single stars.  Second, the
timescale for CV evolution is increased significantly.  The
distribution of CVs as a function of orbital period has traditionally
been treated under the assumption that the time for a CV to cross from
long period to the period gap was much shorter than a Hubble time
because the timescale for angular momentum loss was small.  For
unevolved secondaries, we find that the evolutionary timescale to
reach the period gap can approach the Hubble time; this raises the
intriguing possibility that finite age effects could be significant
for understanding the period distribution of CVs.

Furthermore, there does not appear to be a discontinuity in angular
momentum loss properties at the onset of the fully convective boundary.
We believe that there are a number of independent lines of evidence
that support this claim; the strongest and most direct is the absence
of an abrupt change in the surface rotation velocities as a function
of age at the relevant mass in open clusters.  

We note that including
angular momentum loss from secondary stars also improves the agreement
between observation and theory for the shortest period CVs (1.3 hr vs 1.1 hr); the
empirical angular momentum loss rate that we infer (1.5 times
gravitational radiation) is at the level Patterson (1998) found was
needed to produce the observed minimum CV period.

Therefore, something other than the angular momentum loss rates must
be responsible for the existence of the CV period gap.  One intriguing
possibility is that the presence of two distinct peaks in the
distribution of CVs could reflect two distinct populations of objects
with different origins.  This is made more likely by the long
timescale for angular momentum loss.  The evolutionary state of the
secondary at the onset of the CV phase and the physics of the common
envelope phase present two potential culprits.

If the primary and secondary have similar mass, the secondary could be
significantly evolved and would therefore obey a different mass-radius
relationship than an unevolved secondary.  A star that is near
hydrogen core exhaustion will be systematically larger than an
unevolved star of the same total mass.  Because the angular
momentum loss timescale is long, there may also be significant nuclear
evolution during the CV phase even for unevolved secondaries that
begin at masses of
order 1 solar mass or higher.
There are a number of long period CVs with radii larger than those
inferred from the unevolved mass-radius relationship, which provides
some support for the idea that some CV secondaries may be of this
type.

Another possible mechanism for producing evolved secondary stars is
related to the physics of the common envelope phase.  The most
powerful argument against a large fraction of evolved secondaries has
been the shape of the IMF, which would predict many more low mass
companions.  However, we note that the angular momentum loss rate will
also have strong consequences during the common envelope phase of
evolution.  Models for common envelope evolution (de Kool 1990; Iben
\& Livio, 1993; Yungelson et al. 1993) show that common envelope
systems with a white dwarf and a main sequence secondary result in
binaries with initial periods significantly longer than CVs: typically
more than $\sim$ 3 days).  The period evolution for such binaries is
very slow with the empirical angular momentum loss rate.  This is
especially true for lower mass secondaries that will have lower loss
rates.  The timescale for the formation of CVs can be longer than age
of the Galaxy.  Only systems with very short orbital periods will go
into contact in a Hubble time (less than $\sim$ 2-5 days, with the
lower bound applicable for the lowest mass stars while the upper bound
is applicable for the highest mass stars). Because the nuclear
timescale for stars above a solar mass can be less than the timescale
for common envelope products to spiral together, this would create a
selection effect favoring evolved higher mass secondary stars as the
companions to CVs.  Prior to the CV phase the secondary star is
evolving, and once it get into contact, it has a higher radius than a
ZAMS star of the same mass.  This radius remains systematically higher
until the secondary becomes fully convective.  This explains 2 things
simultaneously: the better fit to observational data by evolved stars
and the low space density of CVs (see Patterson 1984, 1998).  To state
things another way, CVs are both longer-lived and more difficult to
produce in the presence of mild angular momentum loss.  There would be
some low mass secondaries that, by chance, were born with short enough
periods to cross the period gap.  We stress that there is no a priori
reason for {\it all} CV secondaries to be evolved, but rather that
{\it some} may be and that this could produce two different
populations with different period distributions. 

Another effect that may prove important in this context is the
behavior of chemically evolved stars as they approach the fully
convective boundary.  Once evolved secondaries reach the fully
convective phase they would become fully mixed and move close to the
unevolved mass-radius relationship.  This would correspond to a
significant change in radius over a small range in age, which would in
turn correspond to a small mass loss rate and a relatively long
evolutionary timescale at this critical point.  In this scenario, the
peak in the CV distribution at 3 hours would correspond to a
population of mostly evolved secondaries.  We note that meridional
circulation mixing of such evolved stars could also cause a transition
to a chemically homogeneous state at short period; we intend to
explore models of this type.  In the two simple trial models of this
type that we ran, the transition to a fully convective state actually
occurred at a range of periods; however, we neglected meridional mixing
and nuclear evolution during the CV phase.

Finally, the time-averaged mass loss rate from the assumption of
marginal contact is significantly less than the claimed observational
rates.  This could imply that CVs have a short duty cycle. However 
, we believe that the observational uncertainties in the empirical
mass loss rates should be carefully examined before drawing strong conclusiond of this type.

\section {Acknowledgment}
M.P. would like to thank P. Eggleton and Mark Wagner for useful discussions.

\clearpage
\end{document}